\documentclass[sigchi, screen, anonymous=false]{acmart}
\usepackage{enumitem}

\AtBeginDocument{%
  }

\setcopyright{acmlicensed}
\copyrightyear{2018}
\acmYear{2018}
\acmDOI{XXXXXXX.XXXXXXX}

\acmJournal{TOG}
\acmVolume{37}
\acmNumber{4}
\acmArticle{111}
\acmMonth{8}

\begin{document}

\title{DIDUP: Dynamic Iterative Development for UI Prototyping}

\author{Jenny Ma}
\authornote{Equal contribution}
\email{jenny.ma@columbia.edu}
\affiliation{%
  \institution{Columbia University}
  \city{New York City}
  \state{NY}
  \country{USA}
}

\author{Karthik Sreedhar}
\authornotemark[1]
\email{ks4190@columbia.edu}
\affiliation{%
  \institution{Columbia University}
  \city{New York City}
  \state{NY}
  \country{USA}
}

\author{Vivian Liu}
\email{vivian@cs.columbia.edu}
\affiliation{%
  \institution{Columbia University}
  \city{New York City}
  \state{NY}
  \country{USA}
}

\author{Sitong Wang}
\email{sw3504@columbia.edu}
\affiliation{%
  \institution{Columbia University}
  \city{New York City}
  \state{NY}
  \country{USA}
}

\author{Pedro Alejandro Perez}
\email{pap2153@columbia.edu}
\affiliation{%
  \institution{Columbia University}
  \city{New York City}
  \state{NY}
  \country{USA}
}

\author{Lydia B. Chilton}
\email{chilton@cs.columbia.edu}
\affiliation{%
  \institution{Columbia University}
  \city{New York City}
  \state{NY}
  \country{USA}
}

\begin{abstract}
Large language models (LLMs) are remarkably good at writing code. A particularly valuable case of human-LLM collaboration is code-based UI prototyping, a method for creating interactive prototypes that allows users to view and fully engage with a user interface. We conduct a formative study of GPT Pilot, a leading LLM-generated code-prototyping system, and find that its inflexibility towards change once development has started leads to weaknesses in failure prevention and dynamic planning; it closely resembles the linear workflow of the waterfall model. We introduce DIDUP, a system for code-based UI prototyping that follows an iterative spiral model, which takes changes and iterations that come up during the development process into account. We propose three novel mechanisms for LLM-generated code-prototyping systems: (1) adaptive planning, where plans should be dynamic and reflect changes during implementation, (2) code injection, where the system should write a minimal amount of code and inject it instead of rewriting code so users have a better mental model of the code evolution, and (3) lightweight state management, a simplified version of source control so users can quickly revert to different working states. Together, this enables users to rapidly develop and iterate on prototypes. 
\end{abstract}

\begin{teaserfigure}
    \centering
    \includegraphics[width=\textwidth]{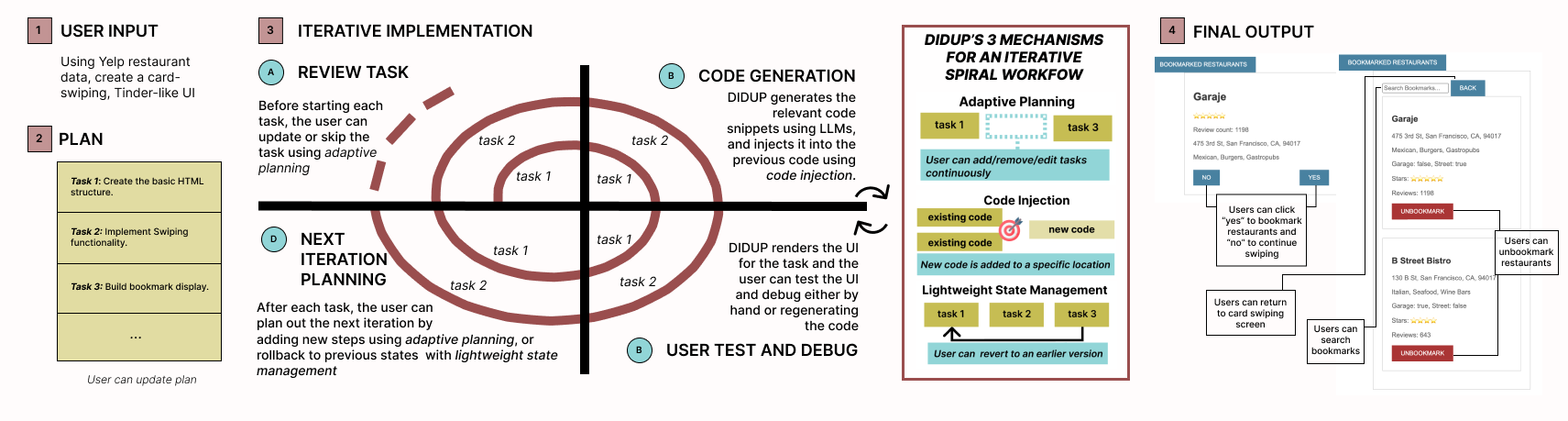}
    \caption{DIDUP is a system for code-based UI prototyping that helps users generate data-driven, interactive UIs in Javascript, CSS, and HTML. The development process uses an iterative spiral model by introducing three mechanisms: adaptive planning, code injection, and lightweight state management. }
    \label{fig:teaser}
\end{teaserfigure}



\maketitle
\begin{figure*}[!htb]
    \centering
    \includegraphics[width=\textwidth]{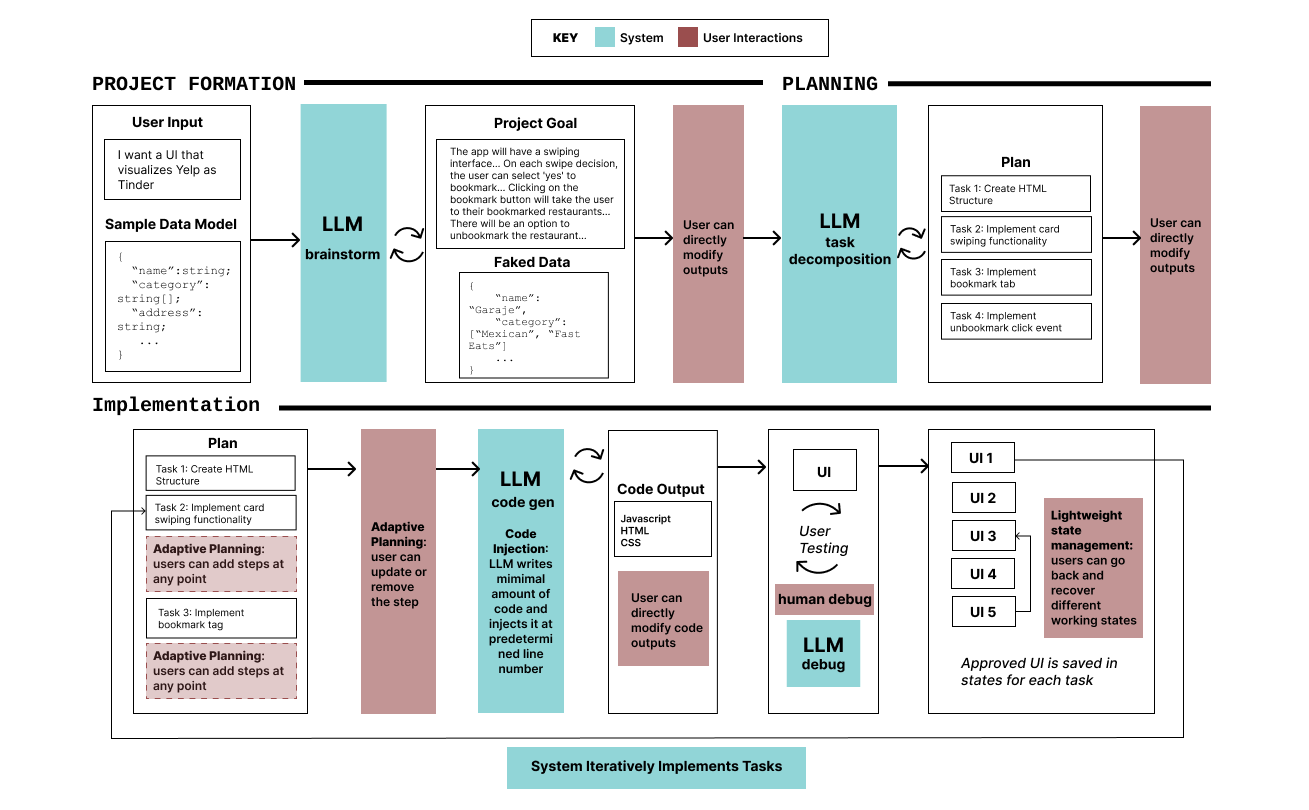}
    \caption{DIDUP system walkthrough. Users begin by inputting their UI goal. DIDUP generates a plan, then implements the tasks iteratively, while keeping the user in control. At every stage, users have the opportunity to approve or change the direction of the code implementation. The implementation stage incorporates adaptive planning, code injection, and lightweight state management in order to create an iterative spiral workflow.}
    \label{fig:system}
\end{figure*}
\section{Introduction}
Large language models (LLMs) are remarkably good at writing code as evidenced by numerous tools including Devin, CoPilot, Gemini, and GPT Pilot \cite{gemini, github_copilot, devin_llm_programmer, gpt_pilot_2024}. Programming is a complex task that requires both high-level, system design planning and low-level understanding of how features can be implemented. Prior work has found that programmers tend to be caught up in the implementation details, debugging rather than thinking at the high-level \cite{Myers}. 
Existing LLM-powered code generation systems show promise in several aspects of human software teams like planning and using multiple agents for specific roles, such as a design architect, product owner, tech lead, and a “code monkey.” 
Some even have infrastructure to evaluate and iterate on LLM-generated code, resulting in more robust outputs. 
These existing systems suggest that LLMs can greatly enhance developer productivity and effectiveness.

A particularly valuable case of human-LLM collaboration is creating code-based user interface (UI) prototypes -- prototypes that enable users to view and fully interact with a UI. 
For example, a user may want to compare how card-swiping interactions differ from a news-feed layout when selecting restaurants to eat at. 
Code-based UI prototypes are valuable for testing data-driven applications that require an interactive component, and often serve as a basis for subsequent versions of the product. In some cases it can even serve as a minimum viable product. 
Whereas there are existing tools for non-code UI prototyping like Figma, and for full stack coding for large code bases like Devin, Gemini and CoPilot, there are currently few tools that support coding for UI prototypes that are both usable and functional \cite{github_copilot, gemini, devin_llm_programmer, figma}. 
It is not a task that can be fully automated, but rather is an iterative and exploratory process where users adapt their designs as they incrementally build and test their prototypes. Automatic code generation can be valuable to alleviate tedious aspects of the development process, but the user must remain in control in order to guide the system.

In software development, there are two main paradigms of writing code: (1) the classic waterfall model, and (2) the iterative spiral model. The waterfall model is a linear workflow where each phase of development must be completed before the next phase begins, and stages are traversed sequentially \cite{royce}. It is simple to realize in practice, but limited because of its inflexibility towards changes based on emerging requirements.
It is therefore not suited for applications with evolving requirements and require cyclical testing such as code-based UI prototyping. 
Conversely, the spiral model is an iterative approach to software development where small features and ideas are continuously designed, implemented, and evaluated as they are discovered \cite{boehm}.
Unlike the waterfall mode, the spiral model supports shifts in project design as developers determine what works and what doesn’t work. 
Prototyping is fundamentally an iterative process, and thus any tool for prototyping should adopt a spiral development process.

\section{Background and Approach}
GPT-Pilot is the leading open-source code prototyping tool powered by LLMs; of the tools we tried, it produced the most working prototypes  \cite{devin_llm_programmer, gpt_pilot_2024}. GPT Pilot has a 4-step workflow: 
first, the user inputs a project goal, and the system asks questions to formalize the specification. Second, GPT Pilot creates an implementation plan, breaking down the project into subtasks. The user is not consulted on this. Third is the execution phase. For each task, GPT Pilot requires human approval before progressing to the next task; if there are bugs it will attempt to debug it. Fourth, after development, there is a documentation phase. While GPT Pilot does have some human input, it is essentially a waterfall model; it lacks sufficient interaction to recover from failures. It also lacks a continuous mechanism that allows the user to elaborate on the prototype and continue improving it. 

We propose three novel mechanisms that can guide the architecture for code-based UI prototyping systems that follow an iterative spiral model: 
\begin{enumerate}[label=\arabic*)]
\item \textbf{Adaptive Planning}: systems must support continual updates in designs and plans based on feedback and implementation. This allows for flexibility in development and for users to elaborate on initial project plans. 
\item \textbf{Code Injection}: when executing sequential tasks, LLMs often rewrite prior working code, which is confusing for users because they don’t know what was removed or added. By injecting the minimal amount of code necessary to a target location, code injection allows for safe code modifications, providing users with a clear mental model of the code’s evolution. 
\item \textbf{Lightweight State Management}: errors inevitably occur, and in situations where the machine cannot fix them, it provides a simplified version of source control. Users can quickly revert to different working states and rapidly prototype explorations. 
 \end{enumerate}

\section{System}

We introduce DIDUP, a web application to assist code-based UI prototypes. It employs a dynamic, iterative approach to development that allows for human direction at every step, as can be seen in \ref{fig:system}. DIDUP is a Flask application that runs on the web, featuring a Python backend and a Typescript frontend. It is designed for users familiar with front-end development who want to prototype faster and helps users author web applications in HTML, CSS, and Javascript.

\begin{figure} [!htb]
    \centering
    \includegraphics[width=\columnwidth]{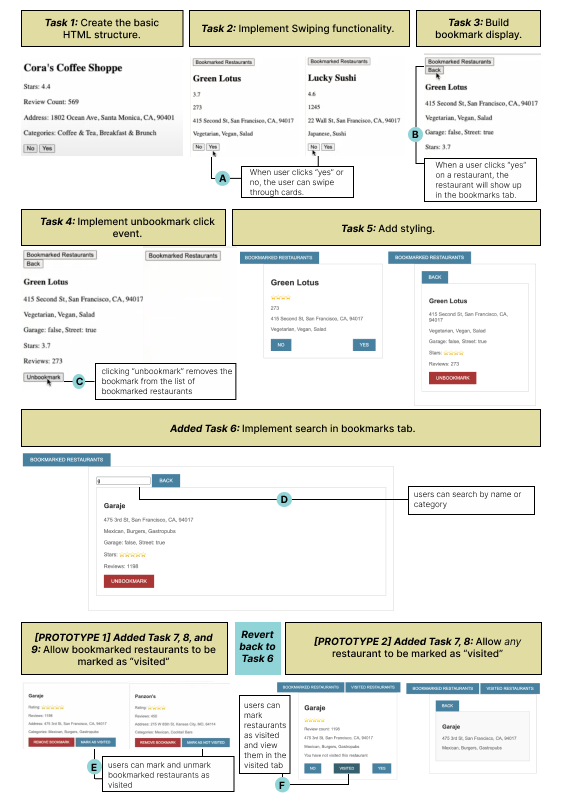}
    \caption{The figure above shows examples of the UI output at each task. DIDUP saves the code for each UI. At each task, the user can test interactions with the interface. The user can also add tasks to help prototype new features and test out more interactions.}.
    \label{fig:output}
\end{figure}

We walk through the system with a motivating example (an example of the UI outputs for each tasks can be seen in \ref{fig:output}). The user first inputs the prototype goal: creating a UI that visualizes Yelp restaurants as a card-swiping UI (Tinder) to help users choose restaurants to eat at. The system outputs a detailed specification of the prototype idea and generates synthetic data to populate the interface. The user can regenerate, approve, or modify the outputs. 

After the user approves the project specification, DIDUP moves on to the planning stage. DIDUP breaks down the project into tasks, where each task is the next-smallest testable iteration of the previous task. In our example, the task list is broken down to: 
\begin{enumerate}[label=\arabic*)]
\item Create the basic HTML structure
\item Implement swiping functionality
\item Build the bookmark display
\item Handle the unbookmark click event
\item Add styling
\end{enumerate}
The user can regenerate, modify, or approve this task list. 

The system executes tasks sequentially. In our example, the system implements task 1) Create the HTML structure. The system generates the initial code, and the UI is rendered on DIDUP for the user to interact with. If there are failures, the user can debug by iteratively prompting the system in the right direction to regenerate code, redoing the task, or debugging the code by hand. 

Once the user confirms that it works, they can proceed to task 2) Implement swiping functionality. The user does not modify the task and prompts the system to generate the code. Instead of rewriting the existing code to implement the task, the system performs \textbf{code injection} -- it writes the minimal amount necessary to complete the task and injects it. DIDUP writes 2 code snippets: 
1) Javascript functionality for swiping cards and 2) event listeners for swipes, and injects it at the appropriate lines in the file. 

Code injection makes it easier for the human to make safe changes to the code. It is similar to a developer reviewing code that a colleague would make when creating Github push requests. It gives the user a mental model of the code as it changes and minimizes the risk of global errors and the inadvertent deletion of good code, ensuring the task remains modularized.

Once the user tests and approves task 2, they can proceed with tasks 3, 4, and 5, to allow users to create, see, and undo bookmarks, and add styling, in similar fashion. DIDUP thus finishes the original plan. 

After using the app with the bookmark feature, the user gets an idea for a new feature - the ability to search within their bookmarks. DIDUP allows the user to add this task as task 6 as a part of \textbf{adaptive planning}, the ability to add, update, or remove tasks continuously during project development. 
It is a crucial component of code-based UI prototyping because it provides the adaptability needed to refine and iterate on design ideas and accommodate changes effectively.
To do this, the user adds a sixth step: 6) Implement search in the bookmarks tab, and the system generates code for this new feature.  

While testing their prototype, the user realizes another feature they want to add - they want to mark a bookmarked restaurant as “visited.” In the future, when they’re looking for a new restaurant to try, they don’t look at the restaurants they've already been to. DIDUP adds a field to the datastore of bookmarked restaurants to have a binary variable for “visited,” and generates code so that the UI has a button on every restaurant “card” in the bookmarked section that the user can mark as visited.

The user tests the prototype and sees the feature is fully functioning, but realizes that they’ve made a mistake - they  want to be able to mark \textit{any restaurant as “visited”}, not just the bookmarked ones. Making this change requires modifying the underlying data representation to take “visited” out of bookmarks and onto all restaurants. Thus, it seems safest to do a rollback to the version of the code at task 6 (with bookmark’s implemented, but no favorites). 

To support rollback, our system introduces \textbf{lightweight state management}. DIDUP saves the state of the UI code at every task to allow users to quickly return to a previous state without remnants of previous changes. The user rolls back to step 6. After the rollback, the user can use adapting planning to execute the new task of allowing users to mark any restaurant as visited, and create a new card list of all visited restaurants. The user can now remember all their past restaurant visits, not just the ones that were bookmarked. 

This process allowed the user to prototype a working UI and add features as they tested prototypes. By incorporating adaptive planning, code injection, and lightweight state management, DIDUP allows users to rapidly iterate on code-based UI prototypes. It streamlines the typically tedious process of switching between writing code, designing features, managing states, and frequent testing.
\section{Evaluation}
We conducted a small user study to evaluate DIDUP’s ability to build prototypes compared to GPT Pilot. We tasked two users with creating specific UIs, using both systems. In both cases, GPT Pilot became fixated on using MongoDB to implement a database for the prototype, which the user was not familiar with and did not want to use. Repeated prompting from the user to not use MongoDB did not work. In one case, the user was able to successfully create their application only after restarting the entire process and specifying that MongoDB should not be used in the initial project description input. The other case was ultimately unsuccessful because GPT Pilot entered an infinite loop during the debugging process, and there was no way to roll back to a previous working state.

When using DIDUP, both users created working UI prototypes easily. Both easily added new features and backtracked to make major adjustments. Furthermore, the prototypes they created were more complete. DIDUP-generated UIs had significantly more grouping and stylistic elements than GPT Pilot-generated UIs. 

\section{Future Work}
In the future, we plan to increase the number of participants and formalize the evaluation to more accurately understand how effectively DIDUP's development process can support users in code-based UI prototyping. Currently, our system is limited to front-end code generation in Javascript, CSS, and HTML; we will expand this to full-stack development. Additionally, we can improve upon adaptive planning mechanisms to allow full regenerations of plans if design directions change, or allow for additions of complex features that add plans, or groups of multiple tasks to the initial design. We can also employ a tree-structure for lightweight version control so users can compare existing prototypes against each other. 

For a user to be able to contribute meaningful feedback or direction in terms of system design, they have to understand the space of design possibilities for the UI and UX of their prototype. They have to know what design patterns they can draw upon to represent their data, what alternatives they have, and which would fit best given the current state of their interface. It would be helpful to surface content-aware UI/UX suggestions for a user to consider as they are deeper in their iterative cycle with the user interface. 
Additionally, there are many usability heuristics people use to check the creation of user interfaces. These can include accessibility guidelines, error prevention mechanisms, and readability or user flow suggestions. 

\section{Conclusion}
This paper explores code-based UI prototyping using an iterative spiral model. We introduced three novel mechanisms to guide the architecture for code-based UI prototyping: (1) \textbf{Adaptive Planning} to support continual updates in designs and plans during implementation, (2) \textbf{Code Injection} to prevent rewrites of code within tasks and prevent global errors, and (3) \textbf{Lightweight State Management} to facillitate rollbacks and rapid UI prototypes. We presented DIDUP, a system that utilizes these mechanisms to create UIs in Javascript, CSS, and HTML. Our initial evaluation demonstrated that DIDUP produced more complete and stylistic UIs compared to a baseline of GPT Pilot-created UIs. Additionally, with DIDUP, users were able to backtrack and prevent errors. Following an iterative spiral workflow is crucial when creating systems for code-based UI prototyping, allowing for dynamic development that prevents failures and handles change.  

\bibliographystyle{ACM-Reference-Format}
\bibliography{sample-base}










\end{document}